# Application of data acquisition methods in the field of scientific research of public policy

Ermin Kuka

Emir Tahirović

*The University of Sarajevo, Faculty of administration, Bosnia and Herzegovina*

Apstract

Modern states, in today's time, are faced with an increasingly wide range of competences, functions and responsibilities in the management of social and state processes. In this sense, they must enact numerous public policy (through the public decision-making process), in order to efficiently and timely solve numerous social problems. On the other hand, public policy also represent a special subdiscipline within political science, within political science. As such, they are given increasing importance and importance in the context of scientific research and scientific approaches. Public policy as a discipline of political science have their own special subject and method of research. A particularly important aspect of the scientific approach to public policy is the aspect of applying research methods as one of the stages and phases of designing scientific research. In this sense, the goal of this research is to present the application of scientific research methods in the field of public policy. Those methods are based on scientific achievements developed within the framework of modern methodology of social sciences. Scientific research methods represent an important functional part of the research project as a model of the scientific research system, predominantly of an empirical character, which is applicable to all types of research. This is precisely what imposes the need to develop a project as a prerequisite for applying scientific methods and conducting scientific research, and therefore for a more complete understanding of public policy. The conclusions that will be reached point to the fact that scientific research of public policy can not be carried out without the creation of a scientific research project as a complex scientific and operational document and the application of appropriate methods and techniques developed within the framework of scientific achievements of modern social science methodology.

Keywords: scientific research, design, public policy, methodology, state

# 1. Introduction

All scientific research is classified according to the subject and research method into: theoretical and empirical. When it comes to public policy research, these researches have the characteristics of both theoretical and empirical (multidisciplinary) research. This is of particular importance, bearing in mind the fact that theoretical and empirical research „they interpenetrate each other and serve each other as arguments. They, interconnected, form the whole process from scientific knowledge of individual and special to scientific knowledge of general and general. They simultaneously use various methods of scientific knowledge and practically connect them into a methodological-methodological system" (Termiz, 2022, p. 270-272).

Scientific (theoretical and empirical) research of public policy is gaining more and more importance in recent times. Numerous scientists and researchers, from different areas of social activities and different scientific disciplines, are increasingly dealing with public policy issues.

# 2. Literature review

In short, public policy are defined as concrete (practical, applied) policy on the ground. Andrew Heywood (2004, p. 738) defines public policy as „formal, that is, formulated decisions of government bodies." By public policy is meant „the use of government or political power in the conduct of public policy, activities aimed at solving specific problems or improving the situation in the community" (Priručnik, 2007, p. 5). Australian political scientist Hal Colebatch (2004, p. 79) means public policy „the process of choice, a succession of stages that reflect a rationalist understanding of individual behavior: first one thinks, then acts, and then checks." American professor of political science Thomas Dye (Howlet and Ramesh, 1995, p. 5-6) defines public policy as „anything the government chooses to do or not do." British political scientist William Jenkins (Howlet and Ramesh, 1995, p. 5-6) defines public policy as „a set of interconnected decisions made by political participants or groups of participants, concerning the selection of goals and means for their achievement, within a specific situation in which the implementation of these decisions should be within the power of the participants who made them." Croatian theoretician of public policy Zdravko Petak (2002, p. 51-62) believes that public policy are „a form of social construction aimed at solving the problems faced by a political community", advocating the thesis that „it is of the utmost importance that the expression of public policy is not exclusively linked to government action."

It is evident that public policy are, in fact, „concrete (practical) policy aimed at solving specific issues and problems that burden a certain society, social community, state. It is, therefore, about politics in action" (Tahirović i Kuka, 2020, p. 21-22). They are the basic mechanism or tool available to states (public authorities) for solving specific issues and problems that require an effective solution. Due to its social and practical application and significance, newer approaches to public policy are based on the issues of research and analysis of public policy.

A particularly important area within public policy is the area of public policy research. Public policy research should be based on scientific principles within the framework of modern methodology of social sciences.

## 3. Research - Methods of scientific research

The scientific research of crime is quite demanding and complex research, with an almost equal application and representation of both theoretical and empirical scientific research. However, since the phenomenon of crime is primarily and primarily an empirical phenomenon, this means that scientific research into crime is predominantly empirical in nature.

Within the modern methodology of social sciences, the classification of scientific research methods into two groups is universally accepted, namely: 1) general scientific (general knowledge, basic) methods and 2) data acquisition and processing methods. General science (general science, basic) principles are those that „can be applied in the research of subjects of all sciences" (Termiz, 2013, p. 161). This is why these methods are prefixed with "general".

The aforementioned classification of scientific methods in the former Yugoslavia was put into use by a Serbian professor Bogdan Šešić (1983) in several of his capital works in the field of methodology of social sciences.

The key criterion of this classification and division of scientific research methods is "comprehensiveness and purpose", although it is evident that this criterion „is not completely consistently applied. For example, not only general scientific methods are generally applicable in science, but they are also basic methods; in addition, non-general scientific methods have the properties of methods of data acquisition and data processing, while some, one would even say all methods of data acquisition are general scientific, because there are simply no other methods of data acquisition. These are observation, experiment, examination, document content analysis and case study..." (Termiz, 2013, p. 161).

In the research of public policy as a social phenomenon, in addition to general scientific (general knowledge, basic) methods of scientific research, methods of obtaining data are used inevitably, even primarily. Methods of obtaining data are „predominantly, but not exclusively, methods by which data are obtained (understood as partial knowledge about the manifestations of the phenomenon, whereby we understand the manifestations of the phenomenon as indicators - indicators of the existence, activity and properties of the phenomenon, its relationships and connections, roles and functions )" (Termiz, 2013, p. 162).

Bearing in mind the fact that public policy are an empirical activity (a phenomenon that manifests itself in concrete social practice), the application of data acquisition methods during such scientific research is inevitable. It has already been emphasized that the scientific research of public policy is in itself a complex research in its scope and character, due to the wide spectrum of manifestations of that phenomenon (manifestation patterns). Therefore, it is necessary to apply methods that will fully and truthfully (accurately) know all the relevant facts and data about the manifestation of that phenomenon in social practice. It is precisely the methods of obtaining data that are „equipped with techniques (procedures and instruments) necessary for obtaining data, but also for their arrangement and processing" (Termiz, 2013, p. 162-163).

Their application in public policy research has an even greater role and importance due to the fact that these methods are relatively adaptable (elastic) and suitable for various types of combination (with each other and with other methods).

According to the frequency and widespread application of data acquisition methods in scientific research of society and social phenomena, including public policy, the following order of these methods can be reliably determined (Termiz, 2022, p. 378): 1) analysis (content) of documents; 2) examination; 3) observation; 4) experiment; 5) case study and 6) biographical method. It should be emphasized that there is not even a full agreement of scientists and researchers that the mentioned methods are indeed methods of obtaining data. However, on this occasion we will not enter into such broader discussions and analyses.

*3.1. Methods and techniques of obtaining data in the field of public policy*

The basic task when conducting scientific research is the discovery, acquisition, processing, analysis and use of relevant data, based on which relevant conclusions are drawn. Obtaining data, in this sense, is a necessary stage in the implementation of any scientific research, theoretical, empirical or combined.

When conducting scientific research, using previously selected indicators, we „discover certain manifestations of reality, record and articulate them, evaluate them, select, classify and assign meaning. Therefore, there are no ready-made data, but we obtain them through our own activities" (Termiz, 2022).

Ivan Grdešić emphasizes that, when conducting policy analysis, data are „the main 'material' of a policy analyst, they are the 'raw material' of his profession" (Grdešić, 2006, p. 33). It is important for a policy analyst to determine the data that is relevant for policy analysis, and the way, methods and techniques that will be used to obtain them.

When it comes to obtaining data for public policy research, almost all scientific methods are used. The most frequently applied methods of obtaining data in the research of social phenomena and processes, including the research of public policy, are:

1. analysis (content) of documents;
2. examination (techniques: interview and survey);
3. observation;
4. experiment;
5. case study;
6. delphi method.

*Analysis (content) of documents* is a scientific method of empirical analysis. A document is „any human creation, the study of which can provide certain information about people and human society. These creations are written documents, material social functional creations, various forms of transformation of nature and natural laws, and socially useful and human artistic creations" (Termiz, 2013, p. 180).

This method implies a systematic, quantitative and qualitative analysis of the obtained data through the analysis of the most diverse forms of human creations (written, pictorial, artistic and similar). The success of document analysis (content) is based on „quality-selected analysis categories. Categories of analysis are what is sought to be found and coded in the content" (Grdešić, 2006, p. 41). Therefore, when using the method of document analysis (content) in scientific research, including public policy research, both quantitative and qualitative analysis (content) of documents are used. This method is particularly used when analyzing normative and regulatory state acts (constitutions, laws, regulations, regulations, strategies, etc.).

*Examination* is a frequently applied method of obtaining data when researching public policy. The survey is „a very complex and favorite method of direct and at the same time indirect collection of data about the social reality and from it itself... The survey is carried out by asking a question, clear and meaningful, to the respondent who voluntarily and consciously answers" (Termiz, 2022, p. 383). This method directly obtains the relevant and requested data from the respondents. Therefore, according to many scientists and public policy researchers, this method has taken the place of an indispensable method in policy analysis. This is more due to the fact that it implies, first of all, the use of the interview technique, „interviewing members of the policy community, policy actors and entrepreneurs, experts, politicians or members of the political and decision-making elite" (Grdešić, 2006, p. 64). It can serve as an important substitute and supplement for already obtained data through the analysis (content) of documents. Just as the selection of relevant documents is important in the analysis (content) of documents, so in the method of examination (interview) the determination of the sample for examination (interview) is very important. The sample (respondents) is, when researching public policy, most often intentionally determined, depending on what and what kind of data is to be obtained. In this sense, special attention should be paid to the wording and selection of questions. This is more due to the fact that when using this method in policy-research, the most and most often interviewing of elites (experts for certain fields and topics) is carried out. As Ivan Grdešić emphasizes, the examination can „illuminate the world of political networks and communities. Policy-relevant research is focused on the real world of political actors, ways of social and political construction of problems, interpretation of interest structures, human experiences, behavior and values" (Grdešić, 2006, p. 69). Survey as a technique of obtaining data has its wide application in social sciences, including political sciences. Survey research provides shorter data, primarily numerical data, which are later statistically analyzed and processed. In this method too, the determination of a relevant and representative sample is extremely important. The majority of scientists and researchers are of the opinion that a sample in the range of 300-400 respondents is a sufficiently representative sample for further conclusions. In the analysis of public policy, polls are „suitable for determining the real needs or experiences of citizens in a policy, related to setting the policy on the agenda, but also for obtaining new information about the behavior of users of a service. Then, polls are good for analyzing public opinion, and in the analysis of public policy, especially in relation to support for certain solutions in the formulation phase. Also, as a method, the survey is suitable for collecting data on issues of the impact of a certain policy and evaluation of the effectiveness of an existing program in their evaluation." (Miošić et al., 2014, p. 41). Therefore, the survey is suitable for obtaining data in all phases of the policy process, which ranks it among the most applied methods of policy analysis. The selection, formulation, and order of questions, as well as the size and complexity of the survey itself (number of questions and comprehensibility) are extremely important in the survey. In this context, the basic features of a good questionnaire are: (Miošić et al., 2014, p. 41):

- comprehensibility and clarity of the question;
- principle: one question – one possible answer;
- brevity of the question;
- clear time frame (next year or past year, NOT "in the future" or "in the past");
- value-free questions;
- non-suggestive questions;
- question filter: in order to determine the actual level of knowledge of the respondents.

The data obtained through the survey „for the needs of the administrative bodies of the state are most often used to explain or continue the started programs. More for the justification of the policy, than for its initial design. Surveys in a policy-making environment can be useful for policy promotion, measuring its success, but they are of little use when specialist knowledge about hard-to-reach and complex policy-problems is needed" (Grdešić, 2006, p. 54-55).

*Observation* is a method that is also applied in public policy research. Observation is „a method of collecting data through direct sensory observation - observation... It is about conscious observation – observation" (Termiz, 2022, p. 378). Policy-analysts and researchers, using this method, in a direct (observation of current phenomena) or indirect (observation of past phenomena based on preserved human creations) way, gain personal insight into the relevant facts and the state of reality, and based on that, obtain the necessary data. Namely, „old solutions to problems can serve as inspiration for developing new ones, more suitable for modern conditions" (Grdešić, 2006, p. 34).

*Experiment* is a method of obtaining data that is used to a lesser extent in public policy research than is the case with other methods. An experiment is „a method in which an experimental situation is created artificially, with an experimental factor, which is purposefully managed and based on the results of which key knowledge is formed" (Termiz, 2022, p. 399). An experiment is „any research in which a certain phenomenon is studied under controlled conditions, regardless of how it was created, naturally or artificially" (Vujević, 1990, p. 83). Thus, the policy-model, as a result of the process of structuring policy problems, enables the policy-analyst to experiment with different types of issues. In this way, relevant data is obtained, which helps the subsequent simpler and easier communication and connection between the policy-analyst and the client of the policy-analysis, and ultimately the audience itself.

*Method Case Study* is used in „research a case - individual, in a series of cases related to each other by subject or several cases related to each other by subject as a whole – mosaic" (Termiz, 2013, p. 182). A case study research each case separately, so the research results are interconnected in order to gain knowledge about the essential properties of the research phenomenon. This method in policy-analysis is applied especially in situations where the preparations for the research are being carried out. Likewise, given that the policy process is very complex and the number of actors and problems (issues) quite broad, individual research of cases can open and facilitate the way to the final conclusion for the policy analyst. Summarizing the conclusions of individual cases leads to a general conclusion or policy proposal for a specific socio-political problem or issue.

**4. Discussion - Data analysis methods**

After obtaining and selecting all the relevant data necessary for carrying out scientific research, including public policy research, there follows the phase of their processing (manual or computer processing, control, arrangement, grouping, classification, presentation, evaluation), and analysis. By the term data processing, we include „all activities on control and arrangement of available data obtained by various scientific methods, their grouping, classification, presentation, as well as evaluation and analysis of data" (Termiz, 2022, p. 469). When it comes to public policy research, special attention is paid to data analysis.

There are several methods of data analysis in public policy research, and the most common are the following:
- representative analysis;
- open coding;
- content analysis;
- descriptive statistics (statistical analysis);
- graphic analysis and presentation of data;
- comparative analysis;
- cost-benefit analysis;
- actor analysis.

*Representational analysis* is one of the simplest methods of data analysis in policy research. This, first of all, is due to the fact that it is a „classical reading during which numbers or words are simply extracted as data... The goal of representational analysis is to reproduce some practice and event as precisely as possible" (Miošić et al., 2014, p. 43). Representational data analysis enables insight into the „field and practice of public policy", considering the fact that „documents are treated as a reflection of a credible representation of reality" (Esmark and Triantafillou, 2007, p. 99-124).

The open coding method is used in situations where it is necessary to compress (condense) a large amount of text material into a smaller amount using codes. Open coding is a procedure that is not previously systematized, prepared, or elaborated, but arises because analysts and researchers in the course of the data analysis process themselves come up with and elaborate codes that make the job of analysis easier for themselves. It is implemented by „first determining the focus of the coding, which can be in the form of a question. For example, if we analyze the policy towards persons with disabilities and its goals, it can be a question: What should the policy towards persons with disabilities in Croatia achieve? We then compress/encrypt the text via three levels of encryption." Thus, „the first level of coding, the codes of the first order, consist of extracting all the quotes from the text (for example, from half a sentence to several sentences) that make up the answer to the question (depict the given perspective). Therefore, we first mark the parts of the text significant for our analysis. The second level of coding, the second-order codes, are obtained by assigning a title to each marked part of the text - a word or phrase that highlights the essence of the quote. Therefore, relevant terms are added to highlighted parts of the text. In the last step, in the third-order codes, all the titles assigned to the quotations are compared and those that are related and related are grouped together. This is how categories are created, on average from 5 to 15 of them, which are the main points of the text material related to our given perspective and question" (Halmi, 2005, p. 214-225).

The method of content analysis has already been elaborated in detail earlier. It also uses coding, but unlike open coding, in content analysis the codes are prepared and systematized in advance. There are two subtypes of content analysis, namely: qualitative and quantitative analysis. Qualitative content analysis shows „whether certain codes exist in the material, and, if so, in what form they exist in the material... Quantitative content analysis focuses on the frequency of occurrence of a certain code in the material - the frequency is measured, that is, how many times something is in text" (Miošić et al., 2014, p. 44).

*Descriptive statistics (statistical analysis)* is one of the most widespread methods of data analysis and processing, including policy research. A policy analyst does not have to be a skilled statistician, but he should have basic knowledge of statistics (understanding of statistical logic), in order to be able to apply the same in policy research. „Precisely due to the fact that political phenomena are massive, very diverse, in most cases suitable for measurement and quantitative expression, statistical procedures and techniques are suitable for application in political research" (Termiz, 2006, p. 7). Descriptive statistics are „of first-class use in the quantitative analysis of political phenomena and policy problems... The results of descriptive statistical procedures can be presented using various graphic, tabular and cartographic techniques" (Grdešić, 2006, p. 73). Descriptive statistics implies the implementation of procedures for grouping, measuring and presenting data, in order to understand their meaning as clearly and simply as possible.

There are three basic types of procedures within descriptive statistics that describe or measure an individual phenomenon (Grdešić, 2006, p. 73):

1. type of measures - what is the nature of the data of the observed phenomenon;
2. measures of central tendency – arithmetic mean, median, mode;
3. dispersion measures – forms of data distribution.

Descriptive statistical analysis is mostly applied when expressing percentage values, based on the processing of survey questionnaires. An example of showing the percentage values of answers from survey questions: „Respondents of the survey data collection, members of representative bodies at all levels of government, in the policy towards people with disabilities, also show a transformation of attitudes about goals according to the social model. When asked about the purposes of state intervention - *What is the main reason why the individual and the community should especially deal with the rights of PWDs* - the members of the representative bodies dominantly think that this guarantees them their human rights as well as any other being (43.7%) and that they, like all other citizens, have the constitutional right to a dignified life (28%). Thus, as many as 71.7% of the members of the representative bodies 'belong' to the human rights model in defining policy goals for PWDs. A total of 23.8% of them 'decided' for the value of solidarity ('because citizens should help each other' - 12.8%; 'because there is a natural human need to help others' - 11.3%), and for legal argumentation ('because this is guaranteed by the laws and national strategies adopted by the Croatian Parliament') – 3.7%. Only 0.3% think that the community should deal with disability because PWDs are weak and powerless, with the classical understanding of the need for care within the medical model. Therefore, the members of the Croatian representative bodies are actually very good at recognizing the development of the long-term goals of the policy towards the PWD" (Petek, 2012, p. 202).

Graphical analysis and presentation of data is an extremely significant and important method of data analysis in policy research. Everything that can be expressed in numerical quantities can also be shown in graphic form. In graphic representation, „the basis consists of already formed numerical statements according to which graphic statements are specially constructed using geometric images, figures, points, lines, surfaces and various bodies" (Termiz, 2006, p. 153-166).

Policy researchers are required to, by using graphs, interest policy research clients, and the wider public, in the results that have been reached, especially since they often deal with extremely complex problems and questions. Because, „a successful graph tells us that the analyst has succeeded in bringing the researched variables (groups of data) into a meaningful relationship and that he understands their mutual interaction" (Grdešić, 2006, p. 83). For this purpose, different types of charts are used, such as: pie chart - pie or cake, histogram - columns as values, merged columns and layered charts, scatter chart, line chart - trend picture. It is important to emphasize that every graphic representation should be accompanied by a proper and accurate textual description. In addition, it should have a title, legend, variable names, visibly displayed values and data, and the like.

*Comparative data analysis* is important due to the fact that it enables the identification of previous good practices in other areas, other policy or other countries. Such insights can be very useful in using the obtained data in solving current issues and problems, which puts policy analysts in a situation where they are already one step ahead of the established starting (initial) place.

*Cost-benefit analysis* concerns, above all, financial indicators and relationships. It is applied in public policy research in situations where it is necessary to decide whether to invest financial means and resources in certain public policy or not, in order to solve socio-political (public policy) problems. It tries to show all variables in the form of monetary value. It is also one of the key limitations of this method, given the fact that not all variables can be presented in such a (monetary) form.

*Actor analysis* is one of the most well-known methods of analysis within public policy. This method determines the number, character, scope, structure, properties, competencies, interrelationships, as well as other variables related to the actors of the process of creating and implementing public policy. In this context, this analysis is particularly important during the process of public policy advocacy, in which relevant and accurate knowledge of the actors involved is crucial in initiating procedures and procedures for the advocacy and representation of various public policy.

## 6. Conclusion, limitations and further research

Theoretical and especially empirical scientific researches of public policy require the application of adequate scientific research methodology. In addition to fulfilling the basic methodological requirements as a prerequisite for scientific research and social phenomena, it is also crucial to ensure the means and methods of obtaining relevant data. In this sense, it is necessary to use adequate methods of obtaining data on public policy, in order to be able to draw relevant conclusions, and to enact better and more efficient public policy. Processed methods, as well as data acquisition techniques, are applicable in public policy research, with the fact that some of them are applied on a larger scale, and some on a smaller scale.

The analyzed methods and two techniques (interview and survey) are suitable for obtaining data on the manifestation of the social phenomenon of public policy. From these identifications, the results of individual public policy, depends on what will be done in the future, i.e. whether it will be necessary to enact new or the same public policy in solving specific socio-political problems. These data, in later stages and processes, can be used as true indicators of the success or failure of adopted public policy. Therefore, it is also necessary that the makers of public decisions (public policy) themselves know the presented methods and techniques of obtaining data on public policy, so that in the end efficient and fair public decisions (public policy) are made.

**References**

Colebatch, H., 2004. Policy. Fakultet političkih znanosti Sveučilišta u Zagrebu.

Esmark, A. and Triantafillou, P., 2007. Document Analysis of Network Topography and Network Programmes, in: Bogason, P. and Zølner, M., (ed.). Methods in Democratic Network Governance. Palgrave Macmillian.

Grdešić, I., 2006. Osnove analize javnih politika. Fakultet političkih znanosti Sveučilišta u Zagrebu.

Halmi, A., 2005. Strategije kvalitativnih istraživanja u primijenjenim društvenim znanostima. Naklada Slap.

Heywood, A., 2004. Politika. Clio.

Howlett, M. and Ramesh M., 1995. Studying Public Policy: Policy Cycles and Policy Subsystems. Oxford University Press.

Miošić, N. et al., 2014. Analiza i zagovaranje javnih politika. EDU centar GONG-a/Centar za cjeloživotno obrazovanje Fakulteta političkih znanosti Sveučilišta u Zagrebu.

Petak, Z., 2002. Komparativne javne politike: mogu li se uspoređivati rezultati djelovanja vlada?. Politička misao. 39 (1), 51-62.

Petek, A., 2011. Transformacija politike prema osobama s invaliditetom u Hrvatskoj: analiza ciljeva. Anali Hrvatskog politološkog društva. 7, 101-122.

Petek, A., 2012. Transformacija politike prema osobama s invaliditetom: primjena policy mreža - doktorska disertacija. Fakultet političkih znanosti.

Priručnik za analizu javnih politika: Uvod u proces kreiranja javnih politika na lokalnom nivou. 2007. ALDI – Agencija za lokalne razvojne inicijative.

Šešić, B., 1983. Osnovi logike. Naučna knjiga, Beograd 1983.

Tahirović, E. and Kuka, E., 2020. Osnove javnih politika. Univerzitet u Sarajevu, Fakultet za upravu – pridružena članica.

Termiz, Dž., 2006. Statističke tehnike i postupci u politikološkim istraživanjima. NIK "Grafit".

Termiz, Dž., 2013. Osnovi metodologije socijalne psihologije. Amos Graf.

Termiz, Dž., 2022. Metodologija društvenih nauka – treće izmijenjeno i dopunjeno izdanje. Fakultet političkih nauka Univerziteta u Sarajevu/Međunarodno udruženje metodologa društvenih nauka.

Vujević, M., 1990. Uvođenje u znanstveni rad u području društvenih znanosti. Informator.